\documentclass{elsart}
\usepackage{epsfig}
\usepackage{amsmath}

\usepackage{amssymb}

\begin{document}

\def\myint{\int \!\! d^{ {\scriptscriptstyle D} } \boldsymbol{x} \; }
\begin{frontmatter}

\title{Modulational instability, solitons and periodic waves in a model of
quantum degenerate Boson-Fermion mixtures}

\author{Juan Belmonte-Beitia, V\'{\i}ctor M. P\'erez-Garc\'{\i}a, and Vadym Vekslerchik}
\address{Departamento de Matem\'aticas, E. T. S. de Ingenieros Industriales, \\
Universidad de Castilla-La Mancha 13071 Ciudad Real, Spain}




\begin{abstract}
In this paper we study a system of coupled nonlinear Schr\"odinger equations modelling
a quantum degenerate mixture of bosons and fermions. We analyze the stability of plane waves, give precise conditions for the existence of solitons 
 and write explicit solutions in the form of periodic waves. We also check that the solitons observed previously in numerical simulations 
 of the model correspond exactly to our explicit solutions and see how plane waves destabilize to form periodic waves.
\end{abstract}

\end{frontmatter}

\section{Introduction}

Bose-Einstein condensates (BEC) made of ultracold atomic alkali gases have proven to be a fertile 
field in the last years for the study of nonlinear matter waves. In this context, many types of nonlinear structures have been predicted to exist and/or experimentally observed. 
Between them we can cite dark \cite{dark1,dark2,TW98} and bright \cite{PG98,bright1,bright2} solitons, gap solitons \cite{gap},  domain wall solitons \cite{Cohen1,Cohen2}, dark-bright solitons
 \cite{Anglin1,Anglin2}, stabilized solitons \cite{stabilized1,stabilized2,stabilized3}, shock waves \cite{shocks1,shocks2,shocks3}, etc.

One of the reasons of such fruitful interplay between theory and experiment is the fact that below the transition temperature for Bose-Einstein condensation in which
 all bosons occupy the same quantum state, these atomic quantum gases are very accurately described by the so-called Gross-Pitaevskii equation for the wavefunction 
 describing the colectivity of atoms thus leading to a complex quantum system with a highly predictable dynamics.

Recently, there has been a strong interest on quantum degenerate mixtures of Bosons and Fermions. There are different theoretical descriptions of these 
mixtures depending on the different weight of physical phenomena. For instance when fermions are dominant a model was proposed in Ref. \cite{Wadati} and further studied in \cite{Kono1}. Solitons in the framework of this model have been found. Other models coming from Thomas-Fermi approximations can be more suitable to describe stationary configurations with large numbers of bosons such as the one used in Ref. \cite{Adhikari1}. Finally, in this paper we will study a recently proposed model \cite{karpiuk,salerno,Maruyama} which includes the dynamics of both the bosonic and  fermionic component 
and is appropriate for describing more general situations, including those with smaller and/or comparable atom numbers described by the equations
 \begin{subequations}
 \label{NLSx}
\begin{eqnarray}
i \hbar \frac{\partial \psi_B(\boldsymbol{r},\tau)}{\partial \tau} & = &  \left[-\frac{\hbar^2}{2m_B} \Delta  + V_B 
+ \frac{4\pi\hbar^2a_{BB}}{m_B}|\psi_B(\boldsymbol{r},\tau)|^2\right] \psi_B(\boldsymbol{r},\tau),  \nonumber \\ & & 
 + \frac{2\pi\hbar^2 a_{BF}}{\mu_{BF}} \left(\sum_{j=1}^{N_F} |\phi_{F,j}(\boldsymbol{r},\tau)|^2\right)\psi_B(\boldsymbol{r},\tau) \label{NLS1}
\\
\label{NLS2}
i  \hbar \frac{\partial \phi_{F,j}(\boldsymbol{r},\tau)}{\partial \tau}  &= &  \left[-\frac{\hbar^2}{2m_F} \Delta+
V_F+  \frac{2\pi\hbar^2 a_{BF}}{\mu_{BF}}   |\psi_B(\boldsymbol{r},t)|^2 \right]\phi_{F,j}(\boldsymbol{r},\tau),
\end{eqnarray}
\end{subequations}
where $\phi_{F,j}(\boldsymbol{r},\tau), j=1,...,N_F,$ with $\boldsymbol{r} \in \mathbb{R}^3$ and $\tau$ being the time in physical units are wavefunctions describing each of the $N_F$ fermions, $\psi_B(\boldsymbol{r},\tau)$ is the mean field wavefunction for the bosonic component, $a_{ij},  i,j=B,F$ are the scattering lengths for $s$-wave collisions between the atoms and $\mu_{BF} = m_Bm_F/(m_B+m_F)$. The normalizations are 
$\int_{\mathbb{R}^n} |\phi_{F,j}(\boldsymbol{r},\tau)|^2d\boldsymbol{r} = 1,  \int_{\mathbb{R}^n} |\psi(\boldsymbol{r},\tau)|^2d\boldsymbol{r} = N_B$. Finally, the trapping potential is given by 
\begin{equation}
V_{B,F} = \frac{1}{2}m_{B,F} \sum_{j=1,2,3} \omega_j^2 r_j^2.
\end{equation}
In the situations considered in this paper we will study situations with $\omega_1 \ll \omega_2 = \omega_3 = \omega_{\perp}$. In that limit, corresponding to 
strongly transversely confined systems for which system becomes quasi-one dimensional.  Then we can use a multiscale expansion of the fully three-dimensional model equations \cite{PG98} to obtain a one-dimensional version of 
Eqs. (\ref{NLSx}), which, in a new set of adimensional units given by $\boldsymbol{x} = \boldsymbol{r}/a_s$ and $t = \tau\omega_{\perp}$, reads
\begin{subequations}
\label{NLS1D}
\begin{eqnarray}
		i \partial_{t} \psi & = &
		- \frac{ 1 }{ 2m_B }\; \partial_{xx} \psi + V_B \psi
		+ \left(g_{BB} \left| \psi \right |^{2}+
			g_{BF} \sum_{j=1}^{N_{F}} \left| \phi_j \right|^{2} \right) \psi,
	\\
		i \partial_{t} \phi_{j}&=&
		- \frac{1 }{ 2m_F }\; \partial_{xx} \phi_j + V_F \phi_j
		+g_{BF} \left| \psi \right|^{2} \phi_j,
		\qquad\qquad  j=1,.....,N_{F},
	\end{eqnarray}
	\end{subequations}	
with 
The constants $g_{BB}, g_{BF}$ and $g_{FF}$  in Eq. (\ref{NLS1D}) are given  by the formulae $g_{BB} = 2a_{BB}/a_{s}$, $g_{BF} = 2a_{BF}/a_{s}\alpha$,
$g_{FF} = 2a_{FF}/a_{s}\alpha^{2}$,
where $\alpha = m_{B}/m_{F}$, $a_{s}=\sqrt{\hbar/(m_{B}\omega_{\bot})}$. The new  wavefuncions $\psi(x,t), \phi_j(x,t)$ describe the density variations of the condensate along the less confined dimension ($x$) and have the same normalization as the old ones, i. e.  $\int_{\mathbb{R}}|\phi_j(x,t)|^2 dx = 1,  \int_\mathbb{R} |\psi(x,t)|^2dx = N_B$

In all the numerical simulations to be shown in this paper we have taken $m_B = 87m_p$ and $m_F = 40m_p$ corresponding to a quantum degenerate mixture of 
$^{40}$K and $^{87}$Rb  such as the one studied experimentally in \cite{KRb,FB2}. Also the trapping frequency $\omega_{\perp} = 215$ Hz coming from the same references will be taken as typical of relevant experimental situations.

Eqs. (\ref{NLS1D}) have been studied numerically in Ref. \cite{karpiuk} and the formation of localized structures containing bosons and fermions has been reported
 in the particular case in which the interspecies scattering length $a_{BF}$ is negative, which is the case of the $^{40}$K-$^{87}$Rb mixture \cite{KRb,FB2}.  
 In the simpler model consisting of two interacting bosonic species a negative interspecies scattering length has been proven 
 to be able to support stable solitons \cite{Belmonte,Adhikari}

To our knowledge there have been no theoretical analyses of Eqs. (\ref{NLS1D}) supporting the existence of localized structures.
 In this paper, we study Eqs. (\ref{NLS1D}) analitically and numerically to provide some insight on the formation of nonlinear coherent 
 structures and use it to complement the numerical results of Ref. \cite{karpiuk}
concerning the formation of soliton trains. 
 
\section{Modulational instability}

\subsection{Analytical results}

Let us first consider the case with $V_B = V_F=0$, then Eqs. (\ref{NLS1D}) have trivial solutions with complex arbitrary constant amplitudes $\psi^B_{0}$ and 
	 $\phi^F_{0j}$ of the form
	 \begin{subequations}
	\begin{eqnarray}
		\psi &=& 
		\psi^B_{0} \exp \left[ i \theta^B(x,t)\right], \\
		\phi_j&=&
		\phi^F_{0j} \exp \left[ i \theta^F_j(x,t)\right],
		\qquad\qquad j=1,........,N_{F},
	\end{eqnarray}	
	\end{subequations}
	for which we get, after simple algebraic calculations that	
	\begin{subequations}
	\begin{eqnarray}
		\theta^{B} \left( x,t \right)&=&
		i \left( g_{BB} \left| \psi^B_{0} \right |^{2}
		+g_{BF} \left| \phi^F_{0j} \right|^{2} \right)t,
	\\
		\theta^F_j \left( x,t \right)&=&
		i \left( g_{FB} \left| \phi^F_{0j} \right|^{2} \right)t,
                  \qquad\qquad j=1,......,N_{F},
	\end{eqnarray}
	\end{subequations}
	for arbitrary amplitudes $ \psi^B_{0}, \phi^F_{0j}$. 
	
	In what follows we will study the modulational instability of these plane-wave solutions. This mechanism has been recognized as the main one leading to the formation of localized structures in the framework of similar models \cite{Belmonte,Mod1,Mod1b,Mod1c,Mod1d,Mod2} and even playing a key role in the transition to spatiotemporal chaos in related models with non-conservative terms \cite{Mod3}.
	
	Now, we considerer small amplitude and phase perturbations, of the form
	\begin{subequations}
	\begin{eqnarray}
		\psi^B &=&
		\psi^B_{0}+\epsilon \psi^B_{1}(x,t),
	\\
		\theta^B &=&
		\theta^B_{0}(t)+\epsilon \theta^B_{1}(x,t),
	\\
		\phi_j &=&
		\phi^F_{0j}+\epsilon \phi^F_{1j}(x,t),
		\qquad\qquad j=1,.....,N_{F},
	\\ 
		\theta^F_j &=&
		\theta^F_{0j}(t)+\epsilon \theta^F_{1j}(x,t),
		\qquad\qquad j=1,......,N_{F},
	\end{eqnarray}
	\end{subequations}
	The resulting equations for the perturbations are linear and can be solved easily 
	by transforming them to Fourier space, which is equivalent to 
	looking for harmonic perturbations $\psi^B_{1}$, $\theta^B_{1}$,   
	$\phi^F_{1j}$ and  $\theta^F_{1a}$ , $j=1,......,N_{F}$
	 proportional to $e^{ikx}e^{\Omega(k) t}$. After solving the system, we get
	 the dependence of the growth rate $\Omega(k)$ of the Fourier mode of the perturbation with wavenumber $k$
       which is given by
	\begin{equation}\label{MIk}
		\left( \Omega(k)^{2}-f_{_{F}}(k) \right) \left( \Omega(k)^{2}-f_{_B}(k) \right)=C(k)^{2},
       \end{equation}
       	where 
	\begin{subequations}
	\begin{eqnarray}
		f_{_B}(k)&=&-\left(2g_{BB} \left( \psi^B_{0} \right)^{2}
		+\frac{k^{2}}{2}  \right)\frac{k^2}{2m_B^2},\\ 
		f_{_{F}}(k)&=&-\frac{k^{4}}{4m_F^2},\\
		C^{2}(k)&=&\frac{1}{m_Bm_F} \left( \psi^B_{0} \right)^{2} g_{BF}^{2} k^{4}
		 \sum_{j=1}^{N_{F}} \left( \psi^F_{0j} \right)^{2}.
	\end{eqnarray}
	\end{subequations}
	Perturbations remain bounded when $\text{Re}[\Omega(k)] \leq 0$. The so-called modulational instability 
	 occurs when $\Omega(k)^{2}>0$ for any $k$. The  most unstable wavenumbers are those with small $k$
	 and a simple analysis of Eq. (\ref{MIk}) leads to the condition
	 \begin{equation}
	 	g_{BF}^{2} >0.
	\end{equation}
        Then, for any $g_{BF} \neq 0$, plane waves are modulationally unstable. The physical implications of this condition are that  plane waves tend to destabilize and form domains either with overlapping  ($g_{BF} < 0$) or separated components 
        ($g_{BF}>0$). This is the reason why, in the numerical simulations of Ref. \cite{karpiuk} the broad background component splits into a series of domains which we will identify in the following subsection with soliton solutions of the model equations. 

   The fact that this model does not have modulationally stable solutions is interesting and different from the situation which happens in multicomponent Bose-Einstein condensates in which there is a typically a regime in which coupled plane waves are stable \cite{Cohen2,Belmonte}.

   \subsection{Numerical results}
   
\begin{figure}
        \epsfig{file=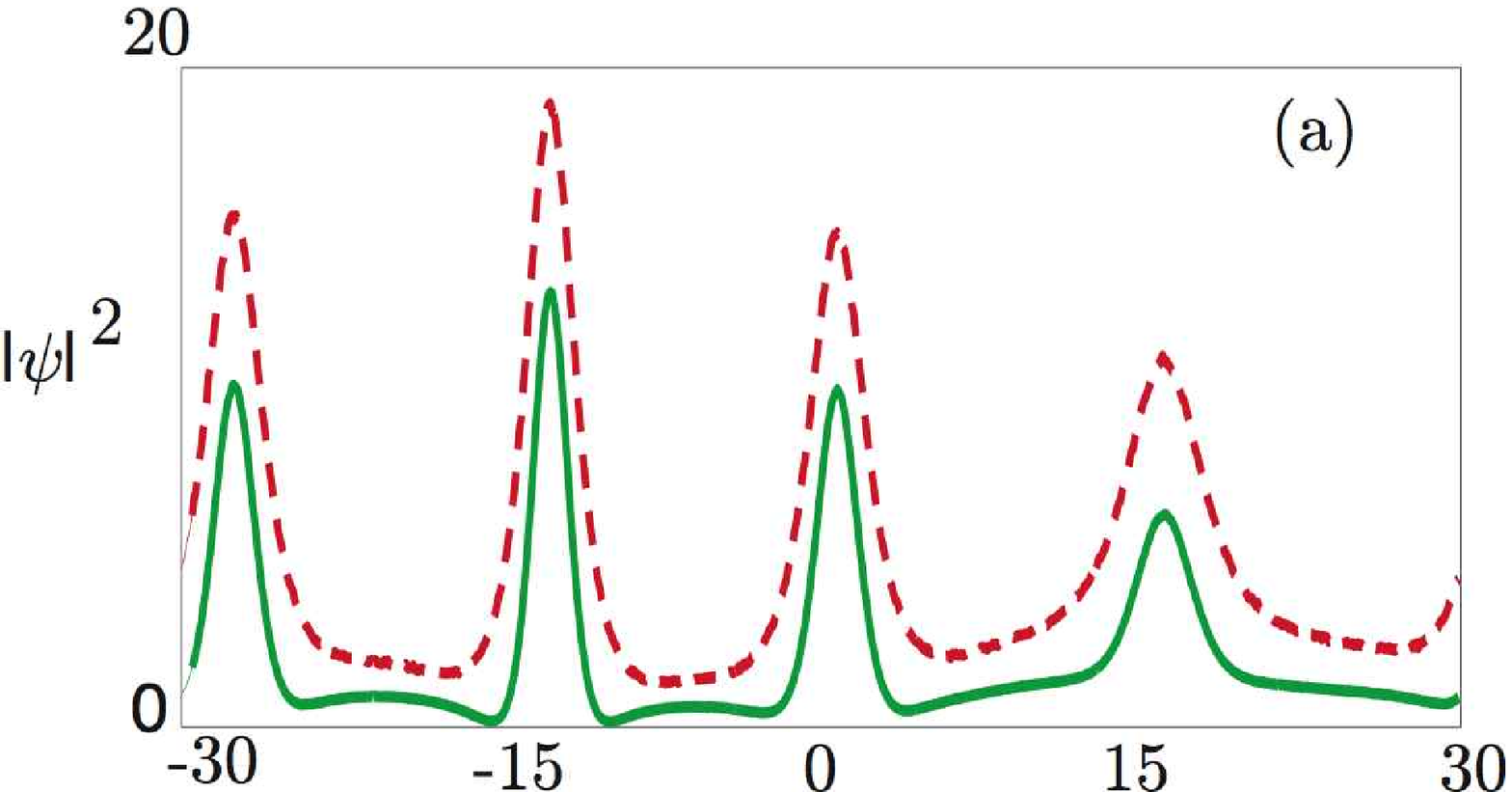,width=7cm}
        \epsfig{file=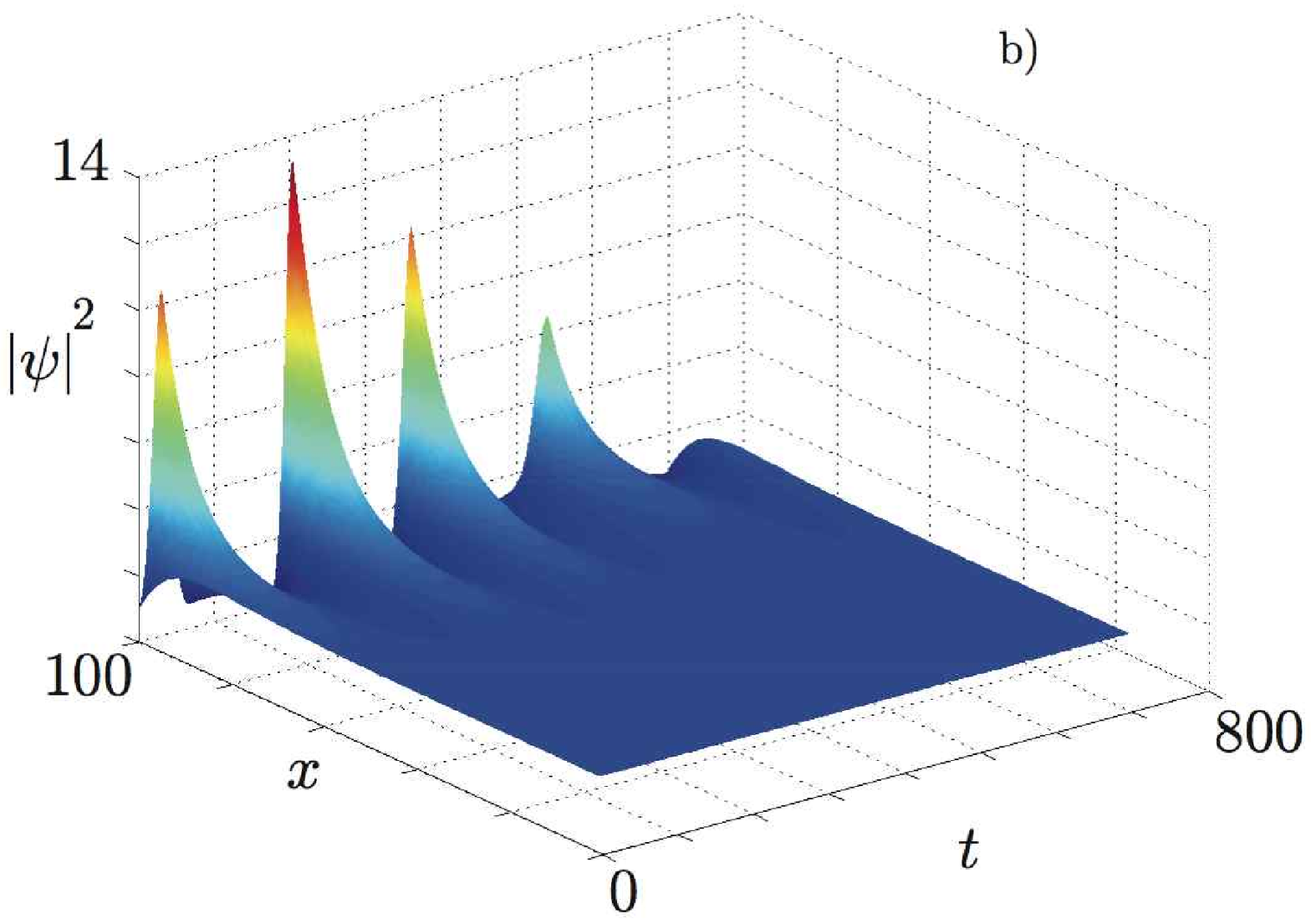, width=7cm}
        \caption{ (a) Formation of trains of localized waves by modulational instability (MI) of plane wave initial data perturbed with a small noise of amplitude 0.015. Parameter values are $g_{BB}=0.0056$ and $g_{BF}=-0.0514$, (corresponding to a physical situation with$a_{BB}=98a_{0},a_{BF}=-410a_{0}$, with $a_{0}$ the Bohr's radius) $N_{B}=300$ and $N_{F}=100$, respectively . The solid green line corresponds to the Bosonic component, whereas the dashed red line identifies the Fermionic component.  (b) 3D surface plot showing the temporal development of the MI into a train of localized wavepackets. \label{prima}}
\end{figure}

   To verify our results and get more insight on the development of the instability we have solved numerically Eqs. (\ref{NLS1D}) with slightly perturbed plane wave initial data with periodic boundary conditions on the spatial domain $x\in [-30,30]$ for different parameter values. We have observe this instability in all of our simulations, which confirms our analytical analysis. An example is shown in Fig. \ref{prima} where it is also manifest a clear tendency to form trains of localized wavepackets (or periodic waves)
       
  To observe the generation of a few solitons we have taken as initial data the ground state confined in a parabolic trap of unit frequency (in our adimensional units) and weak intercomponent interaction. 
In this initial situation we switch off the trapping potential and simultaneously increase the attractive interaction by changing the inter-species scattering length to a value favoring the formation of solitons.
  
\begin{figure}
       \epsfig{file=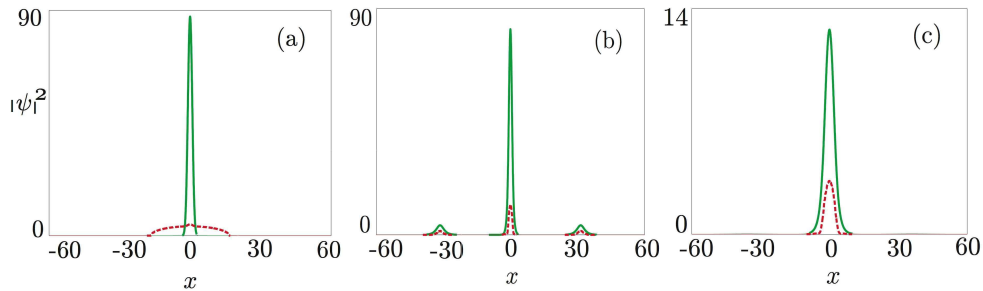,width=14cm}
       \caption{ (a) Density profile of the ground state of a harmonically confined quasi one-dimensional Bose-Fermi mixture for $N_{B}=200$ bosons and $N_{F}=99$ fermions and  $g_{BF}=-0.0038$ (corresponding to an interspecies scattering length of $a_{BF}=-30a_{0}$). (b) Formation of three vector solitons for $g_{BF}=-0.1253$ (corresponding to $a_{BF}=-1000a_{0}$). (c) Formation of a single vector soliton for $g_{BF}=-0.0627$ (for $a_{BF}=-500a_{0}$).  The other parameters are taken as in Fig. \ref{prima} \label{BFef60}}
       \end{figure}

  In Fig. \ref{BFef60} we plot the density profile of the ground state [Fig. \ref{BFef60} (a)] and the outcome after a reduction of  the interspecies scattering length to $a_{BF}=-1000a_{0}$, leading to three localized  structures (solitons), [Fig. \ref{BFef60}. (b)] and to  $a_{BF}=-500a_{0}$ leading to the formation of a single soliton. [see Fig. \ref{BFef60} (c)].
       
\section{Vector soliton solutions}

It is easy to write single-soliton solutions to Eqs. (\ref{NLS1D}). We present a few ones here in the static case, but they can be easily  modified to obtain travelling ones.

First, solutions of the form of coupled bright solitons exist when   $g_{BF} < 0$, and
 $0 < g_{BB} m_B + m_F \left| g_{BF} \right|$, and their explicit forms are
  \begin{subequations}
\label{kar}
\begin{eqnarray}
  \psi & = & 
  A \exp\left( i \omega_B t \right) 
  \frac{ \eta }{ \cosh \eta x }, 
\\
  \phi_j & = & 
  A_j \exp\left( i \omega_F t \right) 
  \frac{ \eta }{ \cosh \eta x},  
\qquad  j=1,...,N_F.
\end{eqnarray}
\end{subequations}
with parameters given by 
\begin{subequations}
\label{etiq}
\begin{eqnarray}
  A^{2} & = & \frac{ 1 }{ m_F \left|g_{BF}\right| },
\\
  \sum_{j=1}^{N_F} A_j^{2} &= &
  \frac{ m_B g_{BB} + m_F \left|g_{BF}\right|  }
       { m_B m_F \left|g_{BF}\right|^{2}}, \\
  \omega_B  & =  & \frac{ \eta^{2} }{ 2m_B }, \ 
  \omega_F  =  \frac{ \eta^{2} }{ 2m_F }.
\end{eqnarray}
\end{subequations}
  This  regime corresponds to the numerical experiments of Ref. \cite{karpiuk}. Our equations \eqref{etiq} provide analytical profiles for the shapes of the solitons found in that paper.  We have checked with our numerical solutions (e.g. those presented in Fig. \ref{BFef60}) that the outcome of the modulational instability process corresponds allways to solitons very well described by our equations. Thus, although one could think of other more general vector soliton solutions such as those discussed in Ref. \cite{Bori2} for two-component systems they do not seem to arise naturally by the MI mechanism in our model.
  
  Although the previous case is probably one of the most interesting situations from the point of view of present experimental capabilities we also write here 
  explicitly the form of  vector solitons solutions for other scenarios for completeness. For instance 
 bright (bosons)-dark (fermions) solutions to Eqs. (\ref{NLS1D}) can be constructed when 
 $ g_{BB} < 0$, $g_{BF} < 0$, $m_F \left| g_{BF} \right| < 
  m_B   \left| g_{BB} \right|$,
 and are given by
\begin{subequations}
\begin{eqnarray}
  \psi & = & 
  A \exp\left( i \omega_B t \right) 
  \frac{ \eta }{ \cosh \eta x }, 
\\
  \phi_j & = & 
  A_j \exp\left( i \omega_j t + i\xi_jx \right) 
  \left[ \xi_j + i\eta \tanh \eta x \right],
\qquad  j=1,...,N_F,
\end{eqnarray}
\end{subequations}
where
\begin{subequations}
\begin{eqnarray}
  A^{2} & = & \frac{ 1 }{ m_F \left|g_{BF}\right| },  \\
  \sum_{j=1}^{N_F} A_j^{2} & =&
  \frac{ m_B \left|g_{BB}\right| - m_F \left|g_{BF}\right| }
       { m_B m_F \left|g_{BF}\right|^{2}  },\\
  \omega_B & =& 
    \frac{ \eta^{2} }{ 2m_B } + 
    \left|g_{BF}\right| 
    \sum_{j=1}^{N_F} 
      \left( \xi_j^{2} + \eta^{2} \right) A_j^{2}, 
\\
  \omega_j & = & - \frac{ \xi_j^{2} }{ 2m_F }
  \qquad  j=1,...,N_F
\end{eqnarray}
\end{subequations}
Other interesting types of vector solitons of Eqs. (\ref{NLS1D}) are coupled dark (bosons)-bright (fermions) solutions, given 
explictly by
\begin{subequations}
\begin{eqnarray}
  \psi & = & 
  A \exp\left( i \omega_B t + i\xi_B x \right) 
  \left[ \xi_B + i\eta \tanh \eta x \right],
\\
  \phi_j & = & 
  A_j \exp\left( i \omega_F t \right) 
  \frac{ \eta }{ \cosh \eta x }, 
  \qquad  j=1,...,N_F.
\end{eqnarray}
\end{subequations}
This solution exists when $ g_{BB} > 0$, $g_{BF} > 0$, and 
$ m_F g_{BF} < m_B g_{BB}$ and the parameters of the solution are given by
 \begin{subequations}
\begin{eqnarray}
  A^{2} & = & \frac{ 1 }{ m_F g_{BF} }, \\
  \sum_{j=1}^{N_F} A_j^{2} & =  &
  \frac{ m_B g_{BB} - m_F g_{BF} }
       { m_B m_F g_{BF}^{2}  },\\
  \omega_B & = &  
    - \frac{ \xi_B^{2} }{ 2m_B } 
    - \frac{ g_{BB} }{ m_F g_{BF} }  
      \left( \xi_B^{2} + \eta^{2} \right), \\
  \omega_F & =& \frac{ \eta^{2} }{ 2m_F }.
\end{eqnarray}
\end{subequations}

Finally, explicit coupled dark soliton solutions having the form
\begin{subequations}
\begin{eqnarray}
  \psi & = & 
  A \exp\left( i \omega_B t + i\xi_B x \right) 
  \left[ \xi_B + i\eta \tanh \eta x \right], 
\\
  \phi_j & = & 
  A_j \exp\left( i \omega_j t + i\xi_jx \right) 
  \left[ \xi_j + i\eta \tanh \eta x \right] 
  \qquad  j=1,...,N_F,
\end{eqnarray}
\end{subequations}
can also be constructed when $0 < g_{BF}$, $m_B g_{BB} < m_F g_{BF}$, the parameters of the solution being
 \begin{subequations}
\begin{eqnarray}
  A^{2} & = & \frac{ 1 }{ m_F g_{BF} }, \\
  \sum_{j=1}^{N_F} A_j^{2} & =&
  \frac{ m_F g_{BF} - m_B g_{BB} }
       { m_B m_F g_{BF}^{2}  },\\
  \omega_B & = & 
    - \frac{ \xi_B^{2} }{ 2m_B }
    - \frac{ g_{BB} }{ m_F g_{BF}} 
      \left( \xi_B^{2} + \eta^{2} \right) 
    - g_{BF} 
      \sum_{j=1}^{N_F} \left( \xi_j^{2} + \eta^{2} \right) A_j^{2}, 
\\
  \omega_j & = & 
    - \frac{ 1 }{ m_F }
      \left( \frac{1}{2} \xi_j^{2} + \xi_B^{2} + \eta^{2} \right), 
  \qquad j=1,...,N_F.
\end{eqnarray}
\end{subequations}
This shows that many different types of vector soliton solutions can be obtained  in the framework of this model. 

\section{Quasiperiodic solutions.}

\subsection{General framework}

The multipeaked structures arising in modulational instability of plane waves can be seen as a soliton train but are most likely to be related to the so called periodic
solutions in the form of Jacobi elliptic functions. This class of solutions arise in many different related nonlinear wave equations \cite{BC1,BC2,BC3} and we will write here some of them for our model equations (\ref{NLS1D}).

We will look for solutions of Eqs. (\ref{NLS1D}) of the form
\begin{subequations}\label{ansatze}
\begin{eqnarray}
  \psi & = & \psi_{1}  \\
  \phi_{j} & = & A_{j} \psi_{2},
  \qquad  j=1,...,N_F,
\end{eqnarray}
\end{subequations}
then, we get
\begin{equation}
  i \partial_{t} \psi_{j} =  
  - \frac{ 1 }{ 2 m_{j} } \; \partial_{xx} \psi_{j}  
  + \left( 
      \sum_{k=1,2} g_{jk} \left| \psi_{k} \right|^{2} 
    \right) \psi_{j},
  \hspace{15mm} j=1,2.
\label{syst-psi-j}
\end{equation}
with $m_1 = m_B, m_2 = m_F$ and
\begin{equation}\label{refect}
 \begin{pmatrix}
 g_{11} & g_{12} \\ g_{21} & g_{22} \end{pmatrix} = 
 \begin{pmatrix}
  g_{BB} &  g_{BF} \sum_{a} \left| A_{a} \right|^{2}  \\  g_{BF} & 0 
  \end{pmatrix}. 
\end{equation}
This means that we are constructing solutions for the boson-fermion coupled system from those of a simpler 
two-component case given by Eqs. (\ref{syst-psi-j}) and (\ref{refect}). In particular we will look for solutions of  the form 
\begin{equation}
  \psi_{j} =  C_{j} \exp\left( i\omega_{j} t \right),
  f_{j}(\xi x) 
\label{psi-j}
\end{equation}
where $f_{j}$ is one of the cnoidal functions $ f_{j} = \text{sn}, \text{cn}, \text{dn}$ satisfying the ODE 
\begin{equation}
  f_{j}'' =  \left( u_{j} + v_{j}f_{j}^{2} \right) f_{j},
\end{equation}
and any two different of them are related by 
\begin{equation}
  p_{1}f_{1}^{2} + p_{2}f_{2}^{2} = 1,
\end{equation}
where $u_{j}$, $v_{j}$ and $p_{j}$ are real numbers.

\subsection{Cases with $f_1 = f_2$.}

First we consider the situation in which $\psi_1$ and $\psi_2$ are proportional, i.e. $f_1 = f_2 \equiv f =  \text{sn}, \text{cn}, \text{dn}$.
Then, after substituting \eqref{psi-j} into Eq. (\ref{syst-psi-j})  one gets that the dispersion relation reads
\begin{equation}
  \omega_{j} =
  \frac{ \xi^{2}u }{ 2m_{j} },
\end{equation}
and the amplitudes are given by the equations
\begin{equation}
  \begin{pmatrix} C_{1}^{2} \\ C_{2}^{2} \end{pmatrix} 
  = 
  \frac{ \xi^{2}v }{ 2 g_{12}g_{21} } 
 \begin{pmatrix} g_{12}/m_{2}  \\
            g_{21}/m_{1} - g_{11}/m_{2}  \end{pmatrix},
\end{equation}
where $u=u_1=u_2$ and $v=v_1=v_2$.
There are different subcases depending on the particular choice of $f$. 
Taking $f = \text{sn}(\xi x, k) $ we get 
\begin{subequations}
\begin{equation}
u = - ( 1 + k^{2} ), v = 2k^{2}.
\end{equation}
 This solution exists if
\begin{equation} 
 g_{BF} > 0, \quad m_B g_{BB}  < m_F g_{BF}.
 \end{equation}
 \end{subequations} 

Next we consider $f = \text{cn}(\xi x, k)$ for which we get 
\begin{subequations}
\begin{equation}
u = 2k^{2} - 1, v = -2k^{2},
\end{equation}
which exists provided 
\begin{equation}
g_{BF} < 0, \quad m_B g_{BB}  > m_F g_{BF}.
\end{equation}
\end{subequations} 
The last possibility with $f_1=f_2$ corresponds to $ f = \text{dn}(\xi x, k) $, its dispersion relation being ruled by 
\begin{subequations}
\begin{equation}
u = 2 - k^{2}, v = -2.
 \end{equation}
This solution exists if
\begin{equation}
 g_{BF} < 0,  \quad m_B g_{BB}  > m_F g_{BF}.
 \end{equation}
 \end{subequations}

\subsection{Cases with $f_1 \neq f_2$.}

In the more general situation $f_1 \neq f_2$ we have a larger variety of combinations.
 In what follows we write the explicit formulae for those situations. First, after substituting  (\ref{psi-j}) into Eq. (\ref{syst-psi-j}) we get 
the dispersion relations,
\begin{subequations}
\begin{equation}
  \omega_{j} = 
  \frac{ \xi^{2} }{ 2m_{j} } 
    \left( u_{j} + \frac{ v_{j} }{ p_{j} } \right) 
  - \frac{ g_{jj} C_{j}^{2} }{ p_{j} }, 
\end{equation}
while the amplitudes are given by 
\begin{equation}
  \begin{pmatrix} C_{1}^{2} \\ C_{2}^{2} \end{pmatrix} 
  = 
  \frac{ \xi^{2} }{ 2 } 
 \begin{pmatrix} 0 & g_{12} p_{1} / p_{2} \\
            g_{21} p_{2} / p_{1} & g_{11} \end{pmatrix}
 \begin{pmatrix} v_{1} / m_{1} \\
            v_{2} / m_{2} \end{pmatrix}.
\end{equation}
\end{subequations}
Now our task is to analyze the different possible combinations of
$\text{sn}$-, $\text{cn}$- and $\text{dn}$-functions.

\subsubsection{Case $ f_{1} = \text{sn}(\xi x, k), \quad f_{2} = \text{cn}(\xi x, k) $.}
The parameters in the dispersion relation are given by
\begin{subequations}
\begin{equation}
  v_{1} = 2k^{2},
  \qquad
  v_{2} = -2k^{2}, 
  \qquad
  p_{1} = p_{2} = 1.
\end{equation}
The amplitudes of the solutions satisfy
\begin{equation}
 \begin{pmatrix} C_{1}^{2} \\C_{2}^{2}  \end{pmatrix} 
  = 
  \frac{ \xi^{2} k^{2} }{ g_{12}g_{21} } 
 \begin{pmatrix} g_{12}/m_{2}  \\
            g_{11}/m_{2} - g_{21}/m_{1}   \end{pmatrix}.
\end{equation}
This solution exists if
\begin{equation}
  g_{BF} > 0,
  \quad
  m_B g_{BB} > m_F g_{BF}.
\end{equation}
\end{subequations}

\subsubsection{Case $ f_{1} = \text{sn}(\xi x, k), \quad f_{2} = \text{dn}(\xi x, k) $.}
The parameters in the dispersion relation are given by
\begin{subequations}
\begin{equation}
  v_{1} = 2k^{2},
  \qquad
  v_{2} = -2, 
  \qquad
  p_{1} = k^{2},
  \quad
  p_{2} = 1.
\end{equation}
The amplitudes of the solutions satisfy
\begin{equation}
 \begin{pmatrix} C_{1}^{2} \\C_{2}^{2} \end{pmatrix} 
  = 
  \frac{ \xi^{2} }{ g_{12}g_{21} } 
 \begin{pmatrix} g_{12}k^{2}/m_{2}  \\
            g_{11}/m_{2} - g_{21}/m_{1} \end{pmatrix}.
\end{equation}
This solution exists if
\begin{equation}
  g_{BF} > 0,
  \quad
  m_B g_{BB} > m_F g_{BF}.
\end{equation}
\end{subequations}

\subsubsection{Case $ f_{1} = \text{cn}(\xi x, k), \quad f_{2} = \text{sn}(\xi x, k) $.}
The parameters in the dispersion relation are given by
\begin{subequations}
\begin{equation}
  v_{1} = -2k^{2}, 
  \quad
  v_{2} = 2k^{2}, 
  \qquad
  p_{1} = p_{2} = 1. 
\end{equation}
The amplitudes of the solutions satisfy
\begin{equation}
 \begin{pmatrix} C_{1}^{2} \\C_{2}^{2} \end{pmatrix} 
  = 
  \frac{ \xi^{2}k^{2} }{ g_{12}g_{21} } 
 \begin{pmatrix} - g_{12}/m_{2}  \\
            - g_{11}/m_{2} + g_{21}/m_{1} \end{pmatrix}
\end{equation}
This solution exists if
\begin{equation}
  g_{BF} < 0,
  \quad
  g_{BB} < 0,
  \quad
  m_B \left| g_{BB} \right| > m_F \left| g_{BF} \right| 
\end{equation}
\end{subequations}

\subsubsection{Case $ f_{1} = \text{cn}(\xi x, k), \quad f_{2} = \text{dn}(\xi x, k) $.}
The parameters in the dispersion relation are given by
\begin{subequations}
\begin{equation}
  v_{1} = -2k^{2},
  \quad
  v_{2} = -2, 
  \qquad
  p_{1} = - \frac{k^{2}}{1-k^{2}}, 
  \quad
  p_{2} = \frac{1}{1-k^{2}}, 
\end{equation}
The amplitudes of the solutions satisfy
\begin{equation}
  \begin{pmatrix} C_{1}^{2} \\C_{2}^{2} \end{pmatrix} 
  = 
  \frac{ \xi^{2} }{ g_{12}g_{21} } 
 \begin{pmatrix} - g_{12}k^{2}/m_{2}  \\
            g_{11}/m_{2} - g_{21}/m_{1}  \end{pmatrix}
\end{equation}
This solution exists if
\begin{equation}
  g_{BF} < 0,
  \quad
  m_B g_{BB} > - m_F \left| g_{BF} \right| 
\end{equation}
\end{subequations}

\subsubsection{Case $ f_{1} = \text{dn}(\xi x, k), \quad f_{2} = \text{sn}(\xi x, k) $.}
The parameters in the dispersion relation are given by
\begin{subequations}
\begin{equation}
  v_{1} = -2, 
  \quad
  v_{2} = 2k^{2}, 
  \qquad
  p_{1} = 1, 
  \quad
  p_{2} = k^{2}. 
\end{equation}
The amplitudes of the solutions satisfy
\begin{equation}
 \begin{pmatrix} C_{1}^{2} \\C_{2}^{2} \end{pmatrix} 
  = 
  \frac{ \xi^{2} }{ g_{12}g_{21} } 
 \begin{pmatrix} - g_{12}/m_{2}  \\
            k^{2}\left( g_{21}/m_{1} - g_{11}/m_{2} \right) \end{pmatrix}
\end{equation}
This solution exists if
\begin{equation}
  g_{BF} < 0,
  \quad
  g_{BB} < 0,
  \quad
  m_B \left| g_{BB} \right| > m_F \left| g_{BF} \right| 
\end{equation}
\end{subequations}

\subsubsection{Case $ f_{1} = \text{dn}(\xi x, k), \quad f_{2} = \text{cn}(\xi x, k) $.}
The parameters in the dispersion relation are given by
\begin{subequations}
\begin{equation}
  v_{1} = -2, 
  \quad
  v_{2} = - 2k^{2},
  \qquad
  p_{1} = \frac{1}{1-k^{2}}, 
  \quad
  p_{2} = - \frac{k^{2}}{1-k^{2}}.
\end{equation}
The amplitudes of the solutions satisfy
\begin{equation}
 \begin{pmatrix} C_{1}^{2} \\C_{2}^{2} \end{pmatrix} 
  = 
  \frac{ \xi^{2} }{ g_{12}g_{21} } 
 \begin{pmatrix} - g_{12}/m_{2}  \\
            k^{2}\left( g_{11}/m_{2} - g_{21}/m_{1} \right) \end{pmatrix}
\end{equation}
This solution exists if
\begin{equation}
  g_{BF} < 0,
  \quad
  m_B g_{BB} > - m_F \left| g_{BF} \right|.
\end{equation}
\end{subequations}

\subsection{Summary}

Table \ref{III} lists the different possible combinations found in the previous subsections 
\begin{table}
\begin{tabular}{|c|c|c|c|}
\hline
\multicolumn{4}{|c|}{$f_1= f_2$:}
\\
\hline
 $g_{BF}$ & 
 $g_{BB}$   &
  $f_{1}$       & 
  $f_{2}$ 
\\[2mm]
\hline
  $g_{BF} > 0$ & 
  $m_B g_{BB} < m_F g_{BF}$ &
  $\text{sn}$ & 
  $\text{sn}$ 
\\
\hline
  $g_{BF} < 0$ & 
  $m_B g_{BB} > m_F g_{BF}$ &
  $\begin{array}{c} \text{cn} \\ \text{dn} \end{array}$ & 
  $\begin{array}{c} \text{cn} \\ \text{dn} \end{array}$ 
\\[2mm]
\hline
\multicolumn{4}{|c|}{$f_1 \neq f_2$:}
\\
\hline
 $g_{BF}$ & 
 $g_{BB}$   &
  $f_{1}$           & 
  $f_{2}$ 
\\[2mm]
\hline
  $g_{BF} > 0$ & 
  $m_B g_{BB} > m_F g_{BF}$ &
  $\begin{array}{c} \text{sn} \\ \text{sn} \end{array}$ & 
  $\begin{array}{c} \text{cn} \\ \text{dn} \end{array}$ 

\\
\hline
  $g_{BF} < 0$ & 
  $g_{BB}<0, \quad  
   m_B \left|g_{BB}\right| > m_F \left|g_{BF}\right| $ &
  $\begin{array}{c} \text{cn} \\ \text{dn} \end{array}$ & 
  $\begin{array}{c} \text{sn} \\ \text{sn} \end{array}$ 
\\
\hline
  $g_{BF} < 0$ & 
  $m_B g_{BB} > - m_F \left|g_{BF}\right| $ &
  $\begin{array}{c} \text{cn} \\ \text{dn} \end{array}$ & 
  $\begin{array}{c} \text{dn} \\ \text{cn} \end{array}$ 
\\
\hline
\end{tabular}
\caption{Summary of exact solutions in form of coupled cnoidal waves to Eqs. (\ref{NLS1D}) with (\ref{ansatze}).  \label{III}}
\end{table}

\section{Conclusions}

In this paper we have studied a mean field model proposed to model the dynamics of Boson-Fermion mixtures.
We have found that plane waves are allways modulationally unstable in this model and have shown by means of numerical simulations 
how their destabilization gives rise to 
structures similar to periodic waves. We have also studied numerically the destabilization of the ground state of the system and the generation of individual solitons.
The explicit form of these solitons have been obtained for the case at hand and for other possible combinations.
Many of these soliton 
solutions of our model equations can be observed in future experiments with Boson-Fermion mixtures. 
Finally, many explicit solutions in the form of periodic waves (sn,cn and dn functions) have been constructed.

\section*{Acknowledgements}

This work has been supported by grants: BFM2003-02832
(Ministerio de Educaci\'on y Ciencia, Spain) and PAI-05-001 (Consejer\'{\i}a de Educaci\'on y Ciencia de la Junta de Comunidades de Castilla-La Mancha, Spain). We want to acknowledge V. Konotop for discussions.

\end{document}